\documentclass[12pt]{iopart}
\usepackage{graphicx}

\usepackage{iopams}
\begin{document}

\title{Shear-dependent Pressure in a Lema\^{i}tre-Tolman-Bondi Metric}

\author{Matts Roos}

\address{Department of Physics, FI-00014 University of Helsinki, Finland}
\ead{matts.roos@helsinki.fi}
\begin{abstract}
We propose to explain the dimming of distant supernovae as the combined effect
of dark energy and a Lema\^{i}tre-Tolman-Bondi (LTB) metric. We take dark energy to have a shear-dependent pressure $p_{DE}=(w_0\, +w_1\, \epsilon)\, \rho_{DE}$
where $\epsilon$ is the ratio of LTB shear to LTB expansion.
\end{abstract}
\maketitle

\section{Introduction}
The observations that distant supernovae of type Ia appear dimmed have stimulated a
vigorous search for models to explain this unexpected fact. Explanations maintaining that
the whole Universe is undergoing an accelerated expansion are certainly very bold when
one recognizes that SNeIa are seen only out to redshifts $z\approx 1.5$. This locally
well-measured part of the Universe is conventionally described by the
Friedmann--Lema\^{i}tre--Robertson--Walker (FLRW)
model for a homogeneous, isotropic and unlimited universe, based on Einstein's equation
in four dimensions. The Einstein tensor $G_{\mu\nu}$ encodes the geometry, and the
stress-energy tensor $T_{\mu\nu}$ encodes the energy density. Thus modifications to
$G_{\mu\nu}$ imply some alternative geometry or metric, whereas modifications to
$T_{\mu\nu}$ involve new forms of energy densities called dark energy.

The traditional explanation of the apparent acceleration is the cosmological constant
$\Lambda$ which can be interpreted either as a modification of the geometry or as a vacuum
energy term in $T_{\mu\nu}$. This fits observational data well, in fact no competing model
fits data better. But the problems with $\Lambda$ are well known: it must be fine-tuned at
some early time to an infinitesimally small value which has no theoretical base,
and it offers no explanation to the coincidence problem.

As a cure one has tried time-dependent cosmological constants (Quintessence models),
scalar fields acting as dark energy (DE), various modifications of Einstein's gravity,
modified metrics, higher dimensional space-times, and spatially inhomogeneous cosmologies.
Gravity may even obey field equations other than Einstein's.

When fitted to observational data no single explanation has been strikingly
successful and all models reproduce the $\Lambda$CDM model quite well, thereby offering
no distinguishable advantage. In this situation we may have to be pessimistic about
the simplicity of the laws of Nature and admit that they could be more complicated
than expected: we may have to introduce more than one modification at a time.
Such many-component models are of course harder to test.

In an attempt of this kind we combined in a previous study \cite{Roos}
the accelerating Chaplygin gas model \cite{Kamenshchik, Bilic} with the
self-decelerating Dvali--Gabadadze--Porrati (DGP) \cite{Dvali, Deffayet}
braneworld model. Interestingly, this two-component model did not approach
the $\Lambda$CDM model in any limit of its parameter space. The DGP deceleration
indeed compensated the accelerating effect of the Chaplygin gas to some extent, but not
enough so that the result did not satisfactorily fit present observational data.

A way to explain the dimming of distant supernovae without DE may be an
inhomogeneous universe described by
the Lema\^{i}tre-Tolman-Bondi solution (LTB) \cite{Lemaitre, Tolman, Bondi} to
Einstein's equation. We could be located near the center of a low-density void
which would distort our measurements of the age of the Universe, the CMB acoustic
scale, and the Baryon Acoustic Oscillations (BAO). LTB models have both their
advocates (e.g. Garc\'{\i}a-Bellido \& Haugb\o lle \cite{Bellido}, Dunsby \&
al. \cite{Dunsby}) and their critics (e.g. Zibin \& al. \cite{Zibin}). Most of
the literature has been cited in the review on observational constraints on
inhomogeneous cosmological models without dark energy by Marra and Notari \cite{Notari}.

In our pessimistic view, future measurements may
show that a modified metric is not enough, so that other components may be needed.
Some recent steps in this sense have been taken by Lasky and Bolejko \cite{Lasky},  Marra and Paakkonen \cite{Marra1},
and Valkenburg \cite{Valkenburg} who studied
two-component models combining the LTB metric with the cosmological constant,
by Marra and Paakkonen \cite{Marra2} who studied an
exact spherically-symmetric inhomogeneous model with $n$ perfect fluids, and
by J. Lee \& al.\cite{Lee} who coupled a Brans-Dicke scalar field to
Horava-Lifshitz gravity.

In the present work we try another two-component model: we combine a
shear-dependent dark energy pressure with the
LTB metric. Needless to say, this model has too much freedom to have any predictive
power at present, but it does add useful attributes to the LTB model.

\section{The model}
The LTB model describes general radially symmetric spacetimes in four dimensions.
The metric can be described by
\begin{equation}
ds^2=-\alpha^2 dt^2+X^2(r,t)\ dr^2+A^2(r,t)\ d\Omega^2,\label{metric}
\end{equation}
where $d\Omega^2=d\theta^2+sin^2\theta\ d\varphi^2$, $A(r,t)$ and $X(r,t)$ are scale functions, and $\alpha(t,r)>0$ is the lapse function. (In a previous version the lapse function was set to $\alpha=1$ which caused the pressure to have zero gradient.)

Assuming a spherically symmetric matter source with baryonic+dark matter density $\rho_M$,
dark energy density $\rho_{DE}$, negligible matter pressure $p_M=0$ and
dark energy pressure density $p_{DE}$, the stress-energy tensor is
\begin{equation}
T^{\mu}_{\nu}=p_{DE}\ g^{\mu}_{\nu}+(\rho_M+\rho_{DE}+p_{DE})\ u^{\mu}u_{\nu}.
\end{equation}
Following the derivation of refs. \cite{Bellido} and \cite{Lasky}, the $(0,r)$ components of Einstein's
equations, $G^0_r=0$,
imply
$\frac{\dot{k}(t,r)}{2(1-k(t,r))}+\frac{\alpha'\dot{A}}{\alpha A'}=0$,
with an arbitrary function $k(t,r)$
playing the r\^{o}le of the spatial curvature parameter.
Here an overdot is designating $\partial_t$ and an
apostrophe $\partial_r$.

The $T^0_0$ component of Einstein's equations gives the
Friedmann--Lema\^{i}tre equation in the LTB metric (\ref{metric})
\begin{equation}
\frac{H_T^2}{\alpha^2}+\frac{2H_T H_L}{\alpha^2}+\frac{k}{A^2}+\frac{k'(t,r)}{AA'}=8\pi G\ (\rho_M+\rho_{DE}).\label{FL}
\end{equation}
Here $H_T=\dot{A}/A$ is the transversal Hubble expansion, $H_L=\dot{A'}/A'$ the longitudinal Hubble expansion.

The $T^r_r$ components of Einstein's equations give
\begin{equation}
2\dot{H_T}+3H_T^2+\frac{k}{A^2}-8\pi G\ p_{DE}=0.\label{Fried2}
\end{equation}
Multiplying each term with $A^3\, H_T$ this can be integrated over time to give
\begin{equation}
H_T^2=\ \frac{F(r)}{A^3}-\frac{k(r)}{A^2}-\frac{8\pi G}{A^3}\int dt\ A^3H_T\ p_{DE}.
\label{H2}
\end{equation}
Here $F(r)$ is an arbitrary time-independent function which arose as an
integration constant.

We now assume that dark energy does not interact with matter. The continuity
condition $T^{\mu\nu}_{\ ;\nu}=0$ then gives two separate continuity equations.
The one for matter is
\begin{equation}
\dot{\rho}_M(r,t)+[2H_T(r,t)+H_L(r,t)]\ \rho_M(r,t)=0,
\end{equation}
and can be integrated to give
\begin{equation}
\rho_M(r,t)=\frac{F(r)}{A^2A'}
\end{equation}
The continuity equation for dark energy is
\begin{equation}
\dot{\rho}_{DE}(r,t)+[2H_T(r,t)+H_L(r,t)]\ [\rho_{DE}(r,t)+p_{DE}(r,t)]=0.\label{cont}
\end{equation}

We now have to specify $p_{DE}(r,t)$ for our model. A very strongly negative pressure
like Chaplygin gas is by itself a poor fit to the accelerated expansion. It could
be combined with a decelerating LTB geometry, but that would then not describe a void.

Alternatively, Quintessence with an equation of state
\begin{equation}
p_{DE}(r,t)=(w_0\ +w_1\, a)\ \rho_{DE}(r,t),
\end{equation}
is meaningless in LTB, because a scale $a$ valid transversally as well as longitudinally
cannot be defined.

We now propose a "Quintessence-like" equation of state
\begin{equation}
p_{DE}(r,t)=(w_0\ +w_1\,\epsilon)\ \rho_{DE}(r,t),\label{EOS}
\end{equation}
where $\epsilon$ is the ratio of shear to expansion as defined by by Garc\'{\i}a-Bellido
and Haugb\o lle \cite{Bellido},
\begin{equation}
\epsilon=\frac{H_T-H_L}{2H_T+H_L},
\end{equation}
and $w_0,\ w_1$ are constants. In FLRW universes $\epsilon$ vanishes identically
since $H_T=H_L=H$. The models studied in ref. \cite{Bellido}
define a normalized shear such that $\epsilon=0$ at redshift $z=0$.

Substituting the equation of state (\ref{EOS}) into the continuity equation (\ref{cont})
the latter becomes
\begin{equation}
\frac{\dot{\rho}_{DE}(r,t)}{\rho_{DE}(r,t)}=-(1+w_0)(2H_T+H_L)-w_1(H_T-H_L).
\end{equation}
This can be integrated to yield
\begin{equation}
\ln \rho_{DE}=-(2+2w_0+w_1)\ln A\ -(1+w_0-w_1)\ln A'+\ln Q(r),
\end{equation}
where $\ln Q(r)$ is an integration constant. Thus
\begin{equation}
\rho_{DE}= \frac{Q(r)}{A^{2+2w_0+w_1}A'^{1+w_0-w_1}}=\frac{Q(r)}{(A^2A')^{1+w_0}}
\left(\frac{A'}{A}\right)^{w_1}.
\end{equation}
Then the dark energy term in Eq. (\ref{H2}) becomes
\begin{equation}
-\frac{8\pi G\,Q(r)}{A^3}\int dt\frac{\dot{A}}{A^{2w_0+w_1}A'^{\,1+w_0-w_1}}
(w_0+w_1\epsilon).\label{final}
\end{equation}

\section{Discussion}
This model now contains three arbitrary functions of $r$: $k(r), F(r), Q(R)$ and two
unknown parameters $w_0, w_1$, thus it has very low predictive power. Depending on the parameters $w_0,\ w_1$ the pressure can
either accelerate or decelerate the expansion, and it can even change from acceleration
to deceleration at some critical redshift when $\epsilon= -w_0/w_1$.
For $w_1=0$ dark energy behaves like Quintessence, and for $w_0=-1,w_1=0$ it becomes
the cosmological constant.

Obviously the integral (\ref{final}) can only be evaluated numerically. This is not
motivated to carry out here.
\section*{Acknowledgement}
It is a pleasure to acknowledge inspiration and help from Juan Garc\'{\i}a-Bellido, Universidad Aut\'{o}noma de Madrid.


\section*{References}
\begin{thebibliography}{10}
\bibitem[1]{Roos} Roos, M., 2008 \emph{Phys. Lett.} B, 666, 420
\bibitem[2]{Kamenshchik} Kamenshchik, A., Moschella, U. \&
 Pasquier, V., 2001 \emph{Phys. Lett.} B, 511, 265
\bibitem[3]{Bilic} Bili\'{c}, N., Tupper, G. B. \& Viollier, R. D., 2002
    \emph{preprint} arXiv: astro-ph/0207423
\bibitem[4]{Dvali} Dvali, G. R., Gabadadze \& Porrati, M., 2000 \emph{Phys. Lett.} B, 485,
    208
\bibitem[5]{Deffayet} Deffayet, D., 2001 \emph{Phys. Lett.} B, 502, 199;\\
Deffayet, D, Dvali, G. R. \& Gabadadze, 2002 \emph{Phys. Rev.} D, 65, 044023
\bibitem[6]{Lemaitre} Lema\^{i}tre, G., 1933 \emph{Ann. Soc. Sci. Bruxelles}, Ser. 1, 53, 51
\bibitem[7]{Tolman} Tolman, R. C., 1934 \emph{Proc. Natl. Acad. Sci. U.S.A.} 20, 169
\bibitem[8]{Bondi} Bondi, H., 1947 \emph{Mon. Not. R. Astron. Soc} 107, 410
\bibitem[9]{Bellido} Garc\'{\i}a-Bellido, J. \& Haugb\o lle, T., 2008 \emph{J. Cosmol. Astropart. Phys.} 0804, 003
\bibitem[10]{Dunsby} Dunsby, P., Goheer, N., Osano, B. \& Uzan, J.-Ph., 2010 \emph{preprint} arXiv:1002.2397 [astro-ph.CO]
\bibitem[11]{Zibin} Zibin, J. P., Moss, A. \& Scott, D., 2008 \emph{Phys. Rev. Lett.} 101,251303
\bibitem[12]{Lasky} Lasky, P. D. \& Bolejko, K., Class. Quantum Grav. (2010) 27, 035011
\bibitem[13]{Notari} Marra, V. \& Notari, A., 2011 \emph{Class. Quantum Grav.} 28, 164004
\bibitem[14]{Marra1} Marra, V. \& Paakkonen, M., 2010, \emph{J. Cosmol. Astropart. Phys.}
12,021
\bibitem[15]{Valkenburg} Valkenburg, W., 2011 \emph{Gen. Rel. Grav.} 44, 2449-2476
\bibitem[16]{Marra2} Marra, V. \& Paakkonen, M., 2012 \emph{JCAP} 01, 025
\bibitem[17]{Lee} Lee, J., Lee, T. H. \& Oh, Ph., 2011 \emph{Phys. Lett.} B, 701,393
\end{thebibliography}
\end{document}